

\titlepage

\Ref\BIPZ {E. Brezin, C. Itzykson, G. Parisi and J.B. Zuber, Comm. Math.
Phys. 59 (1978) 35; E. Brezin, V.A. Kazakov and Al.B. Zamolodchikov, Nucl.
Phys. B338 (1990) 673; D. Gross and N. Miljkovich, Nucl. Phys. B238 (1990)
217; P. Ginsparg and J. Zinn-Justin, Phys. Lett. B240 (1990) 333; G.
Parisi, Europhys. Lett. 11 (1990) 595.}
\Ref\SW {A.M. Sengupta and S.R. Wadia, Int. J. Mod. Phys. A6 (1991) 1961;
G. Mandal, A.M. Sengupta and S.R. Wadia, Mod. Phys. Lett. A6 (1991) 1465.}
\Ref\GK {D.J. Gross and I. Klebanov, Nucl. Phys. B352 (1990) 671.}
\Ref\MOORE {G. Moore, Nucl. Phys. B368 (1992) 557.}
\Ref\DDMW {S.R. Das, A. Dhar, G. Mandal and S.R. Wadia, Int. J. Mod. Phys.
A7 (1992) 5165.}
\Ref\DMW {A. Dhar, G. Mandal and S.R. Wadia, Mod. Phys. Lett. A7 (1992) 3129.}
\Ref\WINFA {I. Bakas, Phys. Lett. B228 (1989) 57; Comm. Math. Phys. 134
(1990) 487; A. Bilal, Phys. Lett. B227 (1989) 406; C. Pope, L. Romans and
X. Shen, Phys. Lett. B236 (1990) 173; Nucl. Phys. B339 (1990) 91; Phys.
Lett. B242 (1990) 401.}
\Ref\WINFB {J. Avan and A. Jevicki, Phys. Lett. 226B (1991) 35; Phys.
Lett. 272B (1991) 17; D. Minic, J. Polchinski and Z. Yang, Nucl. Phys.
B362 (1991) 125; G. Moore and N. Seiberg, Int. J. Mod. Phys. A7 (1992)
2601; S.R. Das, A. Dhar, G. Mandal and S.R. Wadia, Int. J. Mod. Phys. A7
(1992) 5165; A. Dhar, G. Mandal and S.R. Wadia, Int. J. Mod. Phys. A8
(1993) 325; Mod. Phys. Lett. A7 (1992) 3129.}
\Ref\WINFC {E. Witten, Nucl. Phys. B373 (1992) 187; I. Klebanov and A.M.
Polyakov, Mod. Phys. Lett. A6 (1991) 3273.}
\Ref\DDMWB {A. Dhar, G. Mandal and S.R. Wadia, Int. J. Mod. Phys. A8
(1993) 325.}
\Ref\IKS {S. Iso, D. Karabali and B. Sakita, preprint CCNY-HEP-92/6.}
\Ref\JS {A. Jevicki and B. Sakita, Nucl. Phys. B165 (1980) 511.}
\Ref\DJ {S.R. Das and A. Jevicki, Mod. Phys. Lett. A5 (1990) 1639.}
\Ref\PATH {A. Dhar, G. Mandal, S.R. Wadia (to appear).}
\Ref\PEREL {A. Perelomov, Generalized Coherent States and Their
Applications, Springer-Verlag.}
\Ref\DDMWA {S.R. Das, A. Dhar, G. Mandal and S.R. Wadia, Mod. Phys. Letts.
A7 (1992) 71.}
\Ref\POL {J. Polchinski, Nucl. Phys. B346 (1990) 253.}
\Ref\SHENK {S. Shenker, in Random Surfaces and Quantum Gravity, Cargese
Proceedings, Eds. O. Alvarez, E. Marinari and P. Windey (Plenum, 1991).}
\Ref\DMWA {A. Dhar, G. Mandal and S.R. Wadia, A Time-dependent Classical
Solution of $c=1$ String Field Theory and Non-perturbative Effects,
TIFR/TH/92-40, to appear in Int. Jour. of Mod. Phys. A.}
\Ref\OVRUT {R. Brustein and B. Ovrut, Penn. U. preprint UPR-524T.}
\Ref\MENDE {J. Lee and P.F. Mende, preprint BROWN-HET-880.}
\Ref\MSW {G. Mandal, A.M. Sengupta and S.R. Wadia, Mod. Phys. Lett. A6
(1991) 1685.}
\Ref\EWITT {E. Witten, Phys. Rev. D44 (1991) 314.}
\Ref\DMWB {A. Dhar, G. Mandal and S.R. Wadia, Mod. Phys. Lett. A7 (1992) 3703.}
\Ref\MS {E. Martinec and S. Shatashvilli, Nucl. Phys. B368 (1992) 338.}
\Ref\SRD {S.R. Das, Mod. Phys. Lett. A8 (1993) 69; EFI preprint, EFI-93-16.}
\Ref\RUSSO {J.G. Russo, Phys. Lett. B300 (1993) 336.}
\Ref\TY {T. Yoneya, Tokyo (Komaba) preprint UT-KOMABA-92-13, hep-th/9211079.}
\Ref\DMWC {A. Dhar, G. Mandal and S.R. Wadia, Wave Propagation in
Stringy Black-Hole, Tata preprint, TIFR/TH/93-05, hep-th/9304072.}

\title{\fourteenbf Non-Perturbative 2-Dimensional String
Theory}\foot{Based on an invited talk delivered at the
1992 Nishinomiya-Yukawa Memorial Symposium on ``Quantum Gravity'',
Nishinomiya City, Japan, 19-20 November 1992.}
\author{Spenta R. Wadia}
\address{Theoretical Physics Group \break Tata Institute of Fundamental
Research \break Homi Bhabha Road, Bombay 400 005, India \break
e-mail: wadia@tifrvax.bitnet}

\abstract{We review some aspects of the non-perturbative formulation of
2-dim. string theory in terms of non-relativistic fermions.  We derive the
bosonization using $W_\infty$ coherent states in the path-integral
formulation.  We discuss the classical limit and indicate the precise
nature of the truncation of the full theory that leads to collective field
theory.  As applications we briefly discuss classical solutions reponsible
for stringy non-perturbative effects and the 2-dim. black-hole.}

\endpage

\noindent {\bf Introduction}:

Two dimensional string theory presents us with an opportunity to discuss
some basic conceptual and technical issues in string theory.  This is
because this theory can be given a non-perturbative definition (in the
continuum limit) and it exhibits a high degree of solvability, by virtue
of the fact that it can be formulated as a theory of non-relativistic
fermions in 1-dim.[\BIPZ],[\SW],[\GK],[\MOORE].

The key organizing principle of the non-perturbative formulation we shall
present is the $W_\infty$ algebra [\DDMW],[\DMW]
and its representation in terms of non-relativistic
fermions.  This fact enables the construction of the non-perturbative
field theory in terms of the boson field [\DMW] $\hat {\cal U}(p,q,t) =
\displaystyle
\int^{+\infty}_{-\infty} dr \psi^+ (q - {\hbar x\over 2}, t) e^{-ipx}
\psi(q + {\hbar x \over 2},t)$,  which carries a linear
representation of
$W_\infty$ and physically corresponds to the phase space fermion density.
This field theory which acheives an exact bosonization of 1-dim.
non-relativistic fermions has several interesting consequences:

\item{a)} In the weak coupling limit, $\hbar = g_{str}
\rightarrow 0$, the $W_\infty$ algebra goes over to $\omega_\infty$ the
algebra of area preserving diffeomorphisms in 2-dim., which is then the
symmetry algebra of the classical theory
[\WINFA],[\WINFB],[\WINFC].  Further the phase
space of the field theory consists of characteristic functions in phase
space leading to a dynamical droplet picture of the fermi fluid
[\DDMW],[\DMW],[\IKS].  If we
further assume that these droplets have boundaries approximately
described by a quadratic equation in phase space, then we obtain the
equation of collective field theory [\JS],[\DJ], as a truncation of the
complete
system of equations of the weak coupling theory.

\item{b)} It exhibits real time classical solutions which are responsible
for the strong (stringy) non-perturbative effects $\sim \exp(- {1 \over
g_{str}})$, as opposed to the expected $\sim \exp(- {1 \over g^2_{str}})$.

\item{c)} There is a redefinition of the field ${\cal U}(p,q,t)$, which
leads to a description of the 2-dim. space-time as a eternal black-hole
with the fermi level being identified with the black-hole mass.

\noindent {\bf Organization of the paper :}

\nobreak
Section 1, defines the theory and discusses the $W_\infty$ algebra and the
bosonization of the non-relativistic fermions.  We do this by the
path-integral method using $W_\infty$ coherent states.
Section 2, is devoted to the weak coupling (low energy) approximation.  We
discuss $\omega_\infty$, the droplet picture and the approximation that
leads to collective field theory.
Section 3, discusses the strong non-perturbative effects, and Section 4 is
devoted to a brief discussion of the 2-dim. black-hole in the fermi fluid
theory.

\bigskip

\noindent 1.  {\bf Fermion Field Theory, $W_\infty$ algebra
and Bosonization}:

\nobreak
It is well known that the $c=1$ matrix model is described by the field
theory of non-relativistic fermions in 1-dim., defined by the action:
$$
\eqalign{&
S = \int^{+\infty}_{-\infty} dt \int^{+\infty}_{-\infty} dx ~\psi^+ (x,t)
\left(i\hbar {\partial \over \partial t} + {\hbar^2 \over 2} \partial^2_x
- {1\over2} V(x)\right) \psi(x,t), \cr &
V(x) = -x^2 + {g_3 \over \sqrt{N}} x^3 + \cdots \cr &
\int^{+\infty}_{-\infty} dx~\psi^+ (x,t) \psi(x,t) = N, ~~~~~N \rightarrow
\infty}
\eqn\one
$$
In the above we have arranged constants in such a way that in the double
scaling limit as $N \rightarrow \infty$, $\hbar$ is to be identified with
the string coupling $g_{str} = {1 \over \mu}$, $\mu$ is the `fermi level',
measured from the top of the potential.

The most general boson operator is the fermion bilocal $\Phi(x,y) =
\psi(x) \psi^+(y)$ (we have suppressed the time dependence), which by
virtue of the fermion anti-commutation relations satisfies the closed
algebra
$$
[\Phi(x,y),\Phi(x',y')] = \Phi(x,y') \delta(y-x') - \Phi(x',y) \delta(x-y')
\eqn\two
$$
This algebra can be equivalently written in terms of
$$
\hat {\tilde u}(\alpha,\beta) = \int^{+\infty}_{-\infty} dx~\psi(x + {1\over2}
\hbar\beta) \psi^+(x - {1\over2} \hbar \beta) e^{i\alpha x}
\eqn\three
$$
and the commutation relations are
$$
[\hat {\tilde u}(\alpha,\beta),\hat {\tilde u}(\alpha',\beta')] = 2i \sin
{\hbar \over
2} (\alpha\beta' - \beta\alpha') \hat{\tilde u}(\alpha + \alpha',\beta+\beta'),
\eqn\four
$$
we shall see that \two\ and \four\ is the $W_\infty$ algebra.

Let us now define $W_\infty$.  For our purposes we define it as the
algebra of hermitian differential operators in 1-dim.  Further restriction
on the class of admissible operators will become clear in the
applications.  It is most conviniently described by introducing a
generating functional $\hat g(\alpha,\beta) = e^{i(\alpha\hat x - \beta
\hat p)}$, where $\hat x$ and $\hat p$ satisfy the usual commutation
relation, $[\hat x,\hat p] = i\hbar$.  We note that an expansion in powers
of $\alpha$ and $\beta$ generates the hermitian differential operators in
1-dim.:
$$
\hat g(\alpha,\beta) = 1 + i(\alpha \hat x - \beta \hat p) -
{1\over2}\left(\alpha^2 \hat x^2 + \beta^2 \hat p^2 - \alpha\beta(\hat x\hat p
+ \hat p \hat x)\right) + \cdots
$$
The commutation relations of the co-efficient operators are succinctly
described by the commutation relations of $\hat g(\alpha,\beta)$:
$$
[\hat g(\alpha,\beta),\hat g(\alpha',\beta')] = 2i \sin {\hbar \over 2}
(\alpha\beta' - \beta\alpha') \hat g(\alpha+\alpha',\beta+\beta')
\eqn\five
$$
which follow from the well known formula: $\hat g(\alpha,\beta) \hat
g(\alpha',\beta') = e^{{i\over 2} \hbar(\alpha\beta' - \beta\alpha')} \hat
g(\alpha + \alpha',\beta+\beta')$.  Hence \two\ and \four\ are
realizations of $W_\infty$ in the field theory of non-relativistic
fermions and this is made more transparent by noting that $\tilde
u(\alpha,\beta,t) = \displaystyle \int^{+\infty}_{-\infty} dx~\psi(x,t)
\hat g(\alpha,\beta) \psi^+(x,t)$.

\bigskip

\noindent {\bf Analogy with spin in a magnetic field}:

\nobreak
The fermion theory expressed in terms of the bilocal operator $\Phi(x,y,t)
= \psi(x,t) \psi^+(y,t)$ has a suggestive equation of motion.  Let us
introduce the operator $\Phi(t)$, defined by the matrix element
$\Phi(x,y,t) = \langle x|\Phi(t)|y\rangle$, then the fermion equation of
motion implies
$$
i\hbar {\partial \over \partial t} \Phi = [h(\hat x,\hat p),\Phi],
{}~~~h(\hat x,\hat p) = {1\over2} \hat p^2 + {1\over2} V(\hat x).
\eqn\six
$$
Equation \two\ and \six\ are equation of a $W_\infty$ `spin' system, with
$h(\hat x,\hat p)$ acting like a external magnetic field.  This suggests
that we use the method of coherent states to construct a path-integral
representation for correlation functions of the operators $\Phi(x,y,t)$.

\bigskip

\noindent $W_\infty$ {\bf coherent states and path integral}[\PATH]:

\nobreak
The vacuum of our theory is easily constructed by filling the fermi sea to
a certain fermi level, which is determined by the number of fermions.
Call it $|F_0\rangle$.  Coherent states are constructed by the action of
$W_\infty$ group elements on $|F_0\rangle$ [\PEREL]
$$
|g(\theta)\rangle = \hat g_\theta|F_0\rangle, ~~~\hat g \in {\cal G}
W_\infty
\eqn\seven
$$
${\cal G} W_\infty$ is the Lie group corresponding to the $W_\infty$
algebra.  There are various equivalent ways of parametrizing $\hat g$.  In
terms of the generators $\Phi(x,y) = \langle x|\Phi|y\rangle$, $\hat g_\theta =
e^{i tr\Phi\hat\theta}$, where $\hat \theta(\hat x,\hat p)$ is a hermitian
operator with matrix elements $\theta(x,y) = \langle x|\hat\theta|y\rangle$
and the trace is defined by $tr\Phi\hat \theta = \int dx~dy~\Phi(x,y)
\theta(y,x)$.  A convinient expression for $\hat \theta(\hat x,\hat p)$ is
in terms of the Weyl ordered operators $\hat W_{rs} = :\left({\hat p +
\hat x \over \sqrt{2}}\right)^r \left({\hat p - \hat x \over
\sqrt{2}}\right)^s:$.  $\hat \theta = \sum \alpha_{rs} W_{rs}$ (:~: stand
for Weyl ordering) defined by $\hat W_{rs} = \int d\alpha d\beta ~\hat
g(\alpha,\beta) \tilde \omega_{rs} (\alpha,\beta)$, where
$\tilde\omega_{rs} (\alpha,\beta) = \int dp dq ~e^{i(\beta p - \alpha q)}
\omega_{rs} (p,q)$, $\omega_{rs} (p,q) = \left({p+q \over
\sqrt{2}}\right)^r \left({p-q \over \sqrt{2}}\right)^s$.  Note that $\hat
W_r \equiv \hat W_{rr} = \left({\hat p^2 - \hat x^2 \over 2}\right)^r$ are
simply powers of the hamiltonian in case $V(\hat \lambda) = {-\hat x^2
\over 2}$ as $N \rightarrow \infty$.  The commuting set of conserved operators
$\{\hat W_r\}$ is a subalgebra of $W_\infty$.  Call it $H$.  Since $\hat
W_r |F_0\rangle \sim |F_0\rangle$, the distinct coherent states are
basically defined by the coset $W_\infty/H$.

Having specified the $W_\infty$ coherent states defined on the fermi
vaccum $|F_0\rangle$, let us explain their significance in evaluating the
fermion
path integral.  Firstly let us note that we are interested in evaluating
correlation functions only involving the bilocal boson operator
$\Phi(x,y,t)$ or some transform of it.  Because of this it is sufficient
to consider intermediate states in the path integral from the linear span of
$\displaystyle\{\prod_i \Phi(x_i,y_i,t)|F_0\rangle\} = {\cal F}$.  These
states form a complete set and give a resolution of the identity.  On the
other hand we can consider the linear span of the set of coherent states,
$\{e^{itr\Phi\hat\theta}|F_0\rangle \equiv |g_\theta\rangle\} = {\cal C}$.
Clearly any element in the linear span of ${\cal F}$ is in the linear span
of ${\cal C}$ and vice versa.  Hence we can consider inserting a
resolution of the identity in terms of the coherent states, even though
they form a over complete set,
$$
\int d\mu(\theta)|g(\theta)\rangle \langle g(\theta)| = 1
\eqn\eight
$$
The derivation of the path integral then rests on the evaluation of the
short time kernal
$$
K_{t+\epsilon,t} = \langle g(\theta_{t+\epsilon,t})|e^{i{\epsilon \over
\hbar}^H}|g(\theta_t)\rangle
\eqn\nine
$$
when $H = \int dx~\psi^\dagger h(\hat x,\hat p)\psi = tr\Phi h$ is the
hamiltonian (one of the $W_\infty$ generators).  Expanding in $\epsilon$
$$
K_{t+\epsilon,t} = \langle g(\theta_{t+\epsilon})|g(\theta_t)\rangle +
i {\epsilon \over \hbar} \langle g(\theta_t)|H|g(\theta_t)\rangle
\eqn\ten
$$
Let us first evaluate the (simpler) 2nd term in (10),
$$
\eqalign{
\langle g(\theta_t)|H|g(\theta_t)\rangle &= \langle F_0|\hat g^{-1}
(\theta_t) tr(\Phi h) \hat g(\theta_t)|F_0\rangle \cr &
= tr (\langle F_0|\hat g^{-1} (\theta_t)\Phi \hat
g(\theta_t)|F_0\rangle h).}
$$
Now by definition,
$$
\left(\hat g^{-1} (\theta_t) \Phi_{xy} \hat g(\theta_t)\right) = \int
(g^{-1}_{xx'} \Phi_{x'y'} g_{y'y})dx' dy'
$$
where $g_{xy} = (e^{i\hat \theta (\hat x,\hat p)})_{xy}$.  Defining
$\phi^0_{xy} = \langle F_0|\Phi_{xy}|F_0\rangle$, we have
$$
\langle g(\theta_t)|H|g(\theta_t)\rangle = tr (g^{-1} \phi^0 gh)
\eqn\eleven
$$
To evaluate the first term in \ten\ :
$$
\eqalign{
\langle g(\theta_{t+\epsilon}|g(\theta_t)\rangle &= \langle F_0|\hat
g(\theta_{t+\epsilon})^{-1} \hat g(\theta_t)|F_0\rangle \cr &
= 1 + i\epsilon \langle F_0|\hat g^{-1} (\theta_t) i\partial_t \hat
g(\theta_t)|F_0\rangle}
$$
Now it can be proved that,
$$
\langle F_0|\hat g^{-1}(\theta_t) \partial_t \hat g(\theta_t)|F_0\rangle =
tr (\phi_0 g^{-1} \partial_t g)
$$
and hence the short time kernal is
$$
K^{[g_{t+\epsilon},g_t]}_{t+\epsilon,t} = \exp {i\epsilon \over \hbar} (tr
\phi_0 g^{-1} i\hbar \partial_t g + tr g^{-1} \phi^0 gh)
$$
and the finite time kernal is
$$
K = \int \prod_t d\mu (g_t) e^{{i\over\hbar} \int dt~(tr
\phi_0 g^{-1} i\hbar \partial_t g + tr g^{-1} \phi^0 gh)}
\eqn\twelve
$$
where $d\mu (g_t) = \displaystyle \prod_{x\not= y} (g^{-1}_t d g_t)_{xy}$
is the measure over the coset $W_\infty/H$.  The path integral \twelve\
had been defined earlier using heuristic arguments [\DDMWA].  Let us
rewrite it in terms of a hermitian field, $\phi(x,y,t) = \langle
x|\hat\phi (\hat x,\hat p,t)|y\rangle$, corresponding to an element of
$W_\infty$.  For simplicity of presentation we assume that time is
periodic $t \in [0,T]$, $g(0) = g(T)$, and consider the 2-dim. disc ${\cal
D} =
[0,T] \times [0,1]$, such that $\partial {\cal D} = [0,T]$.  We extend
$g(t)$ to
the interior of $D$ : $g(t,s)$, such that $g(t,s=0) = g(t)$ and $g(t,s=1)
= 1$.

Now consider the path integral defined by
$$
K = \int \prod_t d\mu(\hat \phi_t) e^{{i\over\hbar} S(\hat \phi)}
\eqn\thirteen
$$
$$
S(\hat\phi) = -\hbar \int_{\cal D} ds~dt~tr~\hat \phi [\partial_t
\hat\phi,\partial_s \hat \phi] + \oint_{\partial {\cal D}} dt~tr~\hat\phi h
\eqn\fourteen
$$
$$
d\mu (\hat\phi) = \delta(tr\hat\phi - N) \prod_{x,y} \delta(\phi^2_{xy} -
\phi_{xy}) \prod_{x,y} d\phi_{xy}
\eqn\fifteen
$$
Using the polar decomposition of a hermitian matrix $\langle
x|\hat\phi|y\rangle = \phi_{xy} = \langle x|\hat g^{-1} V^0 \hat
g|y\rangle$, $\hat g \in {\cal G} W_\infty$ and $V^0$ is a diagonal
operator, $V^0_{xy} = \delta(x-y) V^0_x$ it is easy to show that
$$
S(g,V^0) = \int_{\partial D} dt(tr(V^0 g^{-1} i\hbar \partial_t g)) +
tr(g^{-1} V^0 gh)
\eqn\sixteen
$$
and
$$
d\mu(g,V^0) = \delta\left(\int V^0_x dx - N\right) \prod_x \left[dV^0_x
\delta(V^{0^2}_x - V_x)\right] J(V^0) \prod_{x\not= y} (g^{-1} d g)_{xy}
\eqn\seventeen
$$
$J(V^0)$ is a Jacobian that depends on the eigenvalues of $V^0$.  Hence
\thirteen\ becomes
$$
K = \int \prod_t d\mu(g,V^0) e^{{i\over\hbar} S(g,V^0)}
\eqn\eighteen
$$
The important point is that the $\delta$-functions in \seventeen\ say that
$V^0$ is the same as $\phi^0$ that appears in \twelve\ upto a permutation
of its eigenvalues which are 0 or 1.  The various permutations correspond
to the different ways of filling the fermi sea of this theory of
non-interacting fermions.  Hence
$$
K = \int \prod_t \prod_{x\not= y} (g^{-1} d g)_{xy} e^{i\over\hbar}
\oint_{\partial D} dt\left[tr(\phi_0^{[P]} g^{-1} i\hbar \partial_t g) +
tr(g^{-1} \phi_0^{[P]} g h)\right]
\eqn\ninteen
$$
where $\phi_0^{[P]}$ is a diagonal operator whose eigenvalues are a
permutation of the eigenvalues of $\phi_0$.  These different permutations
correspond to different ways of filling the fermi sea in this theory of
non-interacting fermions.  Just as $\phi_0$ corresponds to the usual fermi
vaccuum $|F_0\rangle$, $\phi_0^{[P]}$ corresponds to another state
$|F_0^{[P]}\rangle$.  These states can be distinguished by their energies:
$E_0^{[P]} = \langle F_0^{[P]} |H|F_0^{[P]}\rangle = tr(\phi_0^{[P]} h)$.
Even here there can be a degeneracy, but $P=1$ is by definition the lowest
energy state and hence non-degenerate.  Hence the complete equivalence of
\thirteen\ and \twelve\ requires us to specify the 1-point function of
$\Phi(x,y,t)$ so that it leads to a minimum of $\langle tr h\Phi\rangle$.

This completes the {\it proof} of the bosonization.  This was earlier done
by us
using the method of co-adjoint orbits of $W_\infty$ [\DMW].  In fact the
$\delta$-functions in the measure in \fifteen\ and \seventeen\ , specify
the co-adjoint
orbit of $W_\infty$ corresponding to the representation in terms of $N$,
non-relativistic fermions.

\bigskip

\noindent {\bf Path integral in terms of phase space fluid
density} :

\nobreak
Consider the expansion of the $W_\infty$ element $\phi(\hat x,\hat p)$,
which enters the path integral \thirteen\ ,
$$
\phi(\hat x,\hat p) = \int d\alpha d\beta \hat g(\alpha,\beta) \tilde
u(\alpha,\beta)
\eqn\twenty
$$
where
$$
\tilde u(\alpha,\beta) = \int {dp \over 2\pi} {dq \over 2\pi} e^{i(p\beta
- q\alpha)} {\cal U}(p,q)
\eqn\twentyone
$$
\twenty\ and \twentyone\ indeed define the Weyl ordering of $\phi(\hat
x,\hat p)$ corresponding to the classical function ${\cal U}(p,q)$.

In order to express the action in terms of ${\cal U}(p,q,t)$, we state a lemma
due to Moyal:

\noindent Lemma (Moyal):

\nobreak
Given 2 classical functions $f_1(p,q)$ and $f_2(p,q)$ and their
corresponding Weyl ordered operators $\hat f_1(\hat x,\hat p)$ and $\hat
f_2(\hat x,\hat p)$, the classical function corresponding to the
commutator $[\hat f_1,\hat f_2]$ is the fourier transform of the Moyal
bracket,
$$
\eqalign{&
[\hat f_1,\hat f_2] = \int d\alpha d\beta \hat g(\alpha,\beta)
\widetilde{\{f_1,f_2\}}_{MB} (\alpha,\beta) \cr &
\{f_1,f_2\}_{MB} = \sin {\hbar \over 2} (\partial_p \partial_{q'} -
\partial_{p'} \partial_q) f_1 (p,q) f_2(p',q')\big|_{p=p',q=q'}}
\eqn\twentytwo
$$
Note that ${1\over\hbar}\{f_1,f_2\}_{MB} = \{f_1,f_2\}_{PB} + 0(\hbar^2)$,
$PB =$ Poisson bracket.  In formula (13) the trace identity $tr[A,B] = 0$
is implicit.  Restricting to such operators is equivalent to requiring
$\int\int dp~dq\{a(p,q),b(p,q)\}_{MB} = 0$, for the corresponding
classical functions.  This can be achieved by requiring the boundary
condition, $a(p,q)$ and $b(p,q)$ are constant as $p,q \rightarrow \infty$.

Using \twenty\ , \twentyone\ and \twentytwo\ we
can easily see that the action \fourteen\ becomes,
$$
\eqalign{
S(u) = &\int_{\cal D} ds dt \int {dp dq \over 2\pi \hbar} {\cal U}(p,q,t,s)
\left[\hbar^2 \left\{\partial_s {\cal U}(p,q,t,s),\partial_t
{\cal U}(p,q,t,s)\right\}_{MB}\right] \cr &
+ \oint_{\partial D} dt \int {dp dq \over 2\pi \hbar} h(p,q) {\cal U}(p,q,t)}
\eqn\twentythree
$$
and the measure \fifteen\ becomes
$$
d\mu ({\cal U}) = \delta\left(\int {dp dq \over 2\pi \hbar} {\cal
U}(p,q,t) - N\right)
\prod_{p,q} \left[\delta(C(p,q,t) d{\cal U}(p,q,t))\right].
$$
This implies the constraints,
$$
\eqalign{
C(p,q,t) &\equiv \cos {\hbar \over 2} (\partial_p \partial_{q'} - \partial_{p'}
\partial_q) {\cal U}(p,q,t) {\cal U}(p',q',t)\big|_{p=p',q=q'} - {\cal
U}(p,q,t) = 0 \cr &
\cong {\cal U}^2(p,q,t) - {\cal U}(p,q,t) + 0(\hbar^2) = 0}
\eqn\twentyfour
$$
$$
\int {dp dq \over 2\pi \hbar} {\cal U}(p,q,t) = N
\eqn\twentyfive
$$
To derive the equation of motion from \twentythree\ we make a variation
$\delta {\cal U}(p,q,t) = \{\epsilon,{\cal U}\}_{MB}$, that preserves the
constraints \twentyfour\ ,\twentyfive\ :
$${\partial \over \partial t} {\cal U}(p,q,t) + \{h,{\cal U}\}_{MB}
(p,q,t) = 0.
$$
This is the `quantum' version of Liouville's equation.  It
is worth mentioning that if $h = {1\over2} (p^2-q^2)$, the Moyal bracket
equals the Poisson bracket and we get
$$
\partial_t {\cal U} + (p\partial_q + q\partial_p){\cal U} = 0
\eqn\twentysix
$$

\bigskip
\bigskip

\noindent {\bf 2. Weak coupling (classical) limit}:

\nobreak
 From \twentyfour we see that as $\hbar \rightarrow 0$ the constraint
simplies: ${\cal U}^2(p,q,t) = {\cal U}(p,q,t)$, implying that the
configurations that
enter the path integral are characteristic functions corresponding to a
regions in phase space.  The only dynamical part of the characteristic
functions are the boundaries of the region and can indeed give a precise
description of this using the classical limit of $W_\infty$ algebra, the
classical canonical transformations in 2-dim.: $\hbar \rightarrow 0$ in
\four\ gives the poisson bracket of $W_\infty$
$$
\{\tilde u(\alpha,\beta),\tilde u(\alpha',\beta')\} = (\alpha\beta' -
\beta\alpha') \tilde u(\alpha + \alpha', \beta + \beta').
\eqn\twentyseven
$$
We refer the reader for details to ref. [\DDMW],[\DMW],[\DDMWB].

The rest of this section is a discussion of collective field theory.  We
show the precise nature of the approximations that need to be performed on
the field theory to reproduce collective field theory [\JS].

Consider defining the moments of ${\cal U}(p,q,t)$,
$$
\eqalign{&
\rho(q,t) \equiv {\tilde \rho(q,t) \over 2\pi\hbar} =
\int^{+\infty}_{-\infty} {dp \over 2\pi\hbar} {\cal U}(p,q,t) \cr &
\pi(q,t) \rho(q,t) = \int^{+\infty}_{-\infty} {dp \over 2\pi\hbar} p
{\cal U}(p,q,t) \cr &
\pi_2(q,t) \rho(q,t) = \int^{+\infty}_{-\infty} {dp \over 2\pi\hbar} p^2
{\cal U}(p,q,t) ~~~~~~~~~{\rm ,etc.}}
\eqn\twentyeight
$$
The equation of motion for the moments $\rho(q,t),\pi(q,t),\pi_2(q,t)$
etc. can be derived from the equation of motion \twentysix\ ,
$$
\eqalign{&
\partial_t\tilde \rho + \partial_q (\tilde \rho \pi) = 0 \cr &
\partial_t \pi = \partial_q \left({\pi^2 \over 2} + {q^2 \over 2} -
\pi_2\right) + {\partial_q \rho \over \rho} (\pi^2 - \pi_2) \cr &
{\rm etc.}}
\eqn\twentynine
$$
Further more the constraint \twentyfour\ implies further relations amount
the moments.  However in the limit $\hbar \rightarrow 0$, the
constraint can be solved by characteristic functions in phase space.  The
classical ground state is described by
$$
u_0(p,q) = \theta \left(\mu - {p^2 - q^2 \over 2}\right) =
\theta\left[(\sqrt{q^2 + 2\mu} - p) (p + \sqrt{q^2+2\mu})\right]
$$
where the curve ${p^2 - q^2 \over 2} = \mu$ defines the fermi surface.
Collective field theory is defined by parametrizing ${\cal U}(p,q,t)$ near $u_0
(p,q)$ by [\POL]
$$
{\cal U}(p,q,t) = \theta\left[(p_+(q,t) - p) (p - p_-(q,t))\right]
\eqn\thirty
$$
$p_+(q,t)$ and $p_-(q,t)$ are such that $|p_\pm (q,t) - \sqrt{q^2+2\mu}|$ is
small.  This parametrization requires regularization at the classical
turning point $q^2 + 2\mu = 0$.  This says that the low energy excitations
of the fermi fluid near the fermi surface are described by a curve which
is at most quadratic in $p$.  This assumption leads to a
relation between the moments of $\rho,\pi,\pi_2$ etc.  In particular
$$
\pi_2 = \pi^2 + {1 \over 12} \tilde\rho^2 + 0(\hbar)
\eqn\thirtyone
$$
Substituting this in \twentynine\ we obtain the equations of collective
field theory:
$$
\eqalign{&
\partial_t \tilde\rho + \partial_q (\pi\tilde\rho) = 0 \cr &
\partial_t \pi + \pi \partial_q \pi = -\partial_q\left(-{q^2 \over 2} +
{\tilde\rho^2 \over 8}\right)}
\eqn\thirtytwo
$$
These are classical hydrodynamic equations for the density $\tilde\rho$
and velocity $\pi$ of a classical fluid in 1-dim.

\bigskip

\noindent 3. {\bf Stringy non-perturbative effects}:

\nobreak
If we call $g_{str}$, the coupling constant of the string theory then
guided by standard field theory we
expect non-perturbative effects would go as $e^{-c/g^2_{str}}$.  However,
Shenker [\SHENK] has made the important observation that in string
theories
there can be non-perturbative effects which go as $e^{-c/g_{str}}$, and
that this is a generic feature of string theory.  Their existence in the
$c=1$ model has been argued on the basis of the underlying fermion picture
and corresponds to the tunneling of a {\it single} eigenvalue.  The
issue however remained whether one can understand and derive
such effects in terms of classical solutions of a string field theory.
This problem was solved in ref. [\DMWA] in the framework of the
non-perturbative formulation we have described in section 1.  Other
proposals to solve this problem in the context of collective field theory
have been given in ref. [\OVRUT] and [\MENDE].  We disagree with their
conclusions.

Let us
briefly describe the solution by solving the classical equations
\twentyfour\ , \twentyfive\ and \twentysix\ .
{\it It is worth emphasizing that an effect which is genuinely quantum
mechanical in terms of a single fermion can be described entirely by a
classical solution of the ${\cal U}(p,q,t)$ theory in real time}.  An
exact solution of the full non-linear problem is presently beyond our
capabilities.  However we can construct an almost exact solution as follows:

Consider
$$
{\cal U}(p,q,t) = u_0(p,q) + u_1(p,q,t)
\eqn\thirtythree
$$
where $\int {dp dq \over 2\pi\hbar} u_0 = N -1$ and $\int {dp dq \over
2\pi\hbar} u_1 = 1$
($N$ is large).
$u_0(p,q)$ is simply the fermion distribution in the ground state,
$$
u_0(p,q) = {1 \over 2\pi} \int^\mu_{-\infty} d\nu \int^{+\infty}_{-\infty}
{d\lambda \over \cosh {\lambda\over2}} e^{i\left[\nu\lambda -
{1\over\hbar} (p^2-q^2)\tanh {\lambda \over 2}\right]}
\eqn\thirtyfour
$$
it satisfies equations \twentyfour\ , and \twentysix .  $u_1(p,q,t)$ is
the distribution corresponding to the gaussian packet of a single electron
above the fermi surface:
$$
u_1 (p,q,t) = 2e^{-{1\over\hbar}\left[(p\cosh t-q\sinh t - p_0)^2 + (-p
\sinh t + q \cosh t - q_0)^2\right]}
\eqn\thirtyfive
$$
$p^2_0 - q^2_0 = 2E_0$, is the classical energy of the electron.  It
satisfies \twentyfour\ and \twentysix\ exactly.  Note that as $\hbar \sim
\mu^{-1} \rightarrow 0$
$$
\eqalign{&
{\cal U}(p,q,t) \rightarrow \theta\left(\mu - {p^2 - q^2 \over 2}\right) + 2\pi
\hbar \delta(p - \bar p(t)) \delta(q - \bar q(t)) \cr &
\bar p(t) = p_0 \cosh t + q_0 \sinh t \cr &
\bar q(t) = p_0 \sinh t + q_0 \cosh t}
\eqn\thirtysix
$$
is the classical trajectory.  We may chose $p_0 > 0$, $q_0 < 0$ and $p_0 <
|q_0|$ so that the mean classical trajectory lies in the left half space
corresponding to negative energy.

Now even though $u = u_0 + u_1$, satisfies the linear equation \twentysix\
exactly, it does not satisfy \twentyfour\ upto the cross term
$$
c_{10} = \cos {\hbar \over 2} \left(\partial_q \partial_{q'} -
\partial_{q'} \partial_p\right) u_0(p,q) u_1(p',q')\big|_{p = p' \atop q =
q'}
\eqn\thirtyseven
$$
Note that as $\hbar \rightarrow 0$ $u_0$ and $u_1$ have zero overlap as
long as $|\mu| > |E_0|$.  Now it can be shown that if we choose $|\mu| \gg
|E_0| \gg 0$, then
$$
c_{10} \sim e^{-|\mu|/\hbar}
$$
On the other hand the `trickle' of the single fermion from the left or
right leads to an amplitude $\sim e^{-|E_0|/\hbar} \gg e^{-|\mu|/\hbar}$.
Let us define this trickle by introducing the operator
$$
\eqalign{&
\tau = N_+ (t = +\infty) - N_+ (t = -\infty) \cr &
N_+ (t) = \int^\infty_0 dq \int^{+\infty}_{-\infty} {\cal U}(p,q,t) =
\int^\infty_0 dq \rho(q,t)}
$$
Then using \thirtythree\ , we can show that [\DMWA]
$$
\tau \sim e^{-|E_0|/\hbar} \sim e^{-|E_0|/g_{str}}
\eqn\thirtyeight
$$
We mention that similar solutions of the classical equation of motion are
responsible SUSY breaking in the Marinari-Parisi model, except that now
the classical equation for ${\cal U}(p,q,t)$ is more complicated and contains
$\hbar$ explicitly.  This is because the hamiltonian is no longer
quadratic in `$q$' and hence the moyal bracket $\not=$ poisson bracket.

\bigskip

\noindent 4. {\bf The 2-dim. Black-hole}:

\nobreak
It is a fortunate circumstance that 2-dim. string theory has a black-hole
classical solution [\MSW],[\EWITT].  On the other hand we have a
non-perturbative
formulation in terms of non-relativistic fermions.  These fermions as we
have seen move in a 2-dim. space-time which is flat.  The space-time of
the collective excitation (tachyon) differs from the space-time of the
fermions in that space is defined in terms of the time of flight variable
and in fact semi-classically it is a half line, i.e. space-time has a
boundary.  On the
other hand the metric of the 2-dim. black-hole can indeed be redefined so
that it is described by a flat space time with a boundary.
It is this fact that provides the clue that it may indeed be
possible to described the matrix model space-time as a black-hole
space-time, provided one makes an appropriate transformation of the
tachyon field.

In the following we discuss the derivation of the hyperbolic transform of
the fermi fluid distribution that represents tachyon propagation in a
black-hole space-time [\DMWB].  There are also other works in this
direction [\MS],[\SRD],[\RUSSO],[\TY].  Consider defining a field
$\phi(p,q,t)$ by the
following transform:
$$
\phi(p,q,t) = \int dp' dq' K(pq|p'q') {\cal U}(p',q',t)
\eqn\thirtynine
$$
The equation of motion \twentysix\ for ${\cal U}(p,q,t)$ leads to the
fundamental
property that ${\cal U}(p,q,t) = {\cal U}(ue^{-t},ve^t)$ i.e. time
evolution is a
lorentz boost in phase space.  If we parametrize $u = Ee^\theta$ and $v =
\pm E e^{-\theta}$, for real numers $\theta$ and $E$, then ${\cal
U}(ue^{-t},ve^t)
= {\cal U}(Ee^{-t+\theta},\pm Ee^{t-\theta})$ and this can be equivalently
written as $u\left({1\over2} \ell n\left|{u\over v}\right| - t, \pm
|uv|\right)$.  This tells us that the role of `time' is essentially played
by ${1\over2} \ell n\left|{u \over v}\right|$ and that we should consider
a time ordered product essentially w.r.t. ${1\over2} \ell n\left|{u\over
v}\right|$.  The implication of this fact is that if we like to
calculate correlation functions of the $\phi$ fields introduced in
\thirtynine\ using the same sence of time as the correlation functions of
the ${\cal U}$ fields, we require the property $\phi(p,q,t) =
\phi(ue^{-t},ve^t)$ or in other words $\phi$ satisfies the equation of
motion $\partial_t \phi + p\partial_q \phi + q\partial_p \phi = 0$.

An immediate consequence of this is that $K(pq|p'q')$ is lorentz invariant:
$$
K(ue^t,ve^{-t}|u'e^{-t},v'e^t) = K(u,v|u'v')
\eqn\fourty
$$
We also require $K$ to be translation invariant in phase space,
$$
K(u,v|u',v') = K(u-u',v-v')
\eqn\fourtyone
$$
\fourty\ and \fourtyone\ imply that
$$
K(u,v|u',v') = K\left((u-u')(v-v')\right)
\eqn\fourtytwo
$$
The transform \thirtynine\ now takes the form (Define $\phi(u,v) \equiv
T(u,v)$)
$$
T(u,v) = \int du' dv' K\left((u - u') (v-v')\right){\cal U}(u',v').
\eqn\fourtythree
$$

\noindent The next step is from hind-sight: We know that the
continuum theory has a black-hole classical solution (it is the unique
solution of the classical equations of motion in the $\sigma$-model
approach). The tachyon equation of motion in the continuum
theory is,
$$
{\cal D} T_c(u,v) \equiv \left[4\left(uv - {\mu\over2}\right)\partial_u
\partial_v + 2(u\partial_u + v\partial_v) + 1\right] T_c(u,v) \approx 0(T^2_c)
\eqn\fourtyfour
$$
$T_c$ is the continuum tachyon.  The r.h.s. is small in the string
coupling.  We pursue the same idea for \fourtythree\ and require that
$$
{\cal D} T(u,v) \sim ({\rm small~in~the ~string~coupling})
\eqn\fourtyfive
$$
Since string perturbation theory in the matrix model is a statement that
the fermion distribution $\langle {\cal U}(p,q,t)\rangle$ must be non-zero only
in the neighbourhood of the fermi surface $p^2 - q^2 = 2\mu$, the kernal must
satisfy
$$
{\cal D} K\left((u-u') (v-v')\right) \sim o\left(u'v' - {\mu \over
2}\right), ~~~{\rm for}~ u'v' \sim {\mu \over 2}.
\eqn\fourtysix
$$
If we call $(u-u') (v-v') = x$, then \fourtysix\ implies the differential
equation
$$
x K'(x) + {1\over2} K(x) = 0
\eqn\fourtyseven
$$
with solutions
$$
K(x) = \Bigg\{\matrix{{k_+ \over x^{1/2}}, ~~~x > 0 \cr {k_- \over
(-x)^{1/2}}, ~~~x < 0}
\eqn\fourtyeight
$$
$k_+$ and $k_-$ are as yet arbitrary constants.  However $x=0$ is a
singular point and in order to treat it, let us regularize $K(x):
{}~k_+(x+\epsilon)^{-1/2}$, $x > 0$ and $K(x) = k_-(-x+\epsilon)^{-1/2}$, $x
< 0$.  We have used a symmetric regularization.  Now continuity at $K(x=0)$
implies that $k_+ = k_- = k$ and hence $K(x) = k|x|^{-1/2}$,
$$
\eqalign{
T(u,v) &= \int du' dv' {{\cal U}(u+u',v+v') \over |u'v'|^{1/2}} \cr &
= \int du' dv' {{\cal U}(u',v') \over |(u-u') (v-v')|^{1/2}}}
\eqn\fourtynine
$$
which is the result of ref. [\DMWB].

There is another way of arriving at the conclusion that in \fourtyeight\
$k_+ = k_-$.  We appeal to the Euclidean black-hole.  It is known that the
euclidean formulation of the matrix model requires an analytic
continuation: $t \rightarrow it$, $p \rightarrow -ip$ and $q \rightarrow
q$.  This makes the Minkowski co-ordinates $u = {p+q \over 2}$ and $v =
{p-q \over 2}$, complex conjugates of each other, and `time' defined by
${1\over2} \ell n|{u\over v}|$ is now periodic.  Our entire discussion
above goes through without change except that $x = (u-u') (v-v')$ is now
positive or zero and the diff. eqn. \fourtyseven\ has a unique solution
$K(x) \sim k x^{-1/2}$.  If we continue back to Minkowski space-time we have
the solution $K(x) = k|x|^{-1/2}$.

Given \fourtythree\ and \fourtyeight\ we can envisage a non-perturbative
treatment of black-hole physics.  In particular we can investigate the
important question of the
quantum fate of the classical singularity.  Indeed we find that the
singularities are only a malady of the semiclassical expansion.  We refer
for details to the original papers [\DMWB],[\DMWC].

\bigskip

\noindent {\bf Acknowledgement}:

\nobreak
I wish to express my deep appreciation to the organizers of the
Nishinomiya-Yukawa Memorial Symposium, especially Professors Ninomiya and
Kikkawa for the opportunity to present this work and also for a wonderful
conference.

\medskip

\refout

\end